\documentclass[aps,twocolumn,superscriptaddress,prb]{revtex4-2}
\usepackage{amsmath}
\usepackage{amssymb}
\usepackage{latexsym}
\usepackage{amsfonts}
\usepackage{epsfig}
\usepackage{psfrag}
\usepackage{graphicx}
\usepackage{color}
\usepackage{hyperref}

\usepackage{graphicx}

\usepackage{amsmath}

\begin{document}
\newcommand{\s}[1]{\bar{#1}}
\newcommand{\C}[3]{\hat{c}_{#1#2#3}}
\newcommand{\Cd}[3]{{\hat{c}_{#1#2#3}}^{\dagger }}
\newcommand{\N}[3]{{\hat{n}_{#1#2#3}}}
\newcommand{\crangle}{\rangle^\mathrm{corr}}
\renewcommand{\c}[2]{\hat{c}_{#1,#2}}
\newcommand{\cd}[2]{\hat{c}_{#1,#2}^\dagger}
\newcommand{\nn}{\nonumber\\}

\newcommand{\rs}[1]{{\color{blue} #1}}

\newcommand{\red}[1]{{\color{red} #1}}


\title{Back-reaction and correlation effects on pre-thermalization 
in Mott-Hubbard systems} 

\author{Friedemann Queisser}

\affiliation{Helmholtz-Zentrum Dresden-Rossendorf, 
Bautzner Landstra{\ss}e 400, 01328 Dresden, Germany,}


\author{Christian Kohlfürst}

\affiliation{Helmholtz-Zentrum Dresden-Rossendorf, 
Bautzner Landstra{\ss}e 400, 01328 Dresden, Germany,}

\author{Ralf Sch\"utzhold}

\affiliation{Helmholtz-Zentrum Dresden-Rossendorf, 
Bautzner Landstra{\ss}e 400, 01328 Dresden, Germany,}

\affiliation{Institut f\"ur Theoretische Physik, 
Technische Universit\"at Dresden, 01062 Dresden, Germany,}

\date{\today}

\begin{abstract}
For the Fermi-Hubbard model in the strongly interacting Mott 
insulator state, we study the pre-thermalization dynamics after 
a quench. 
To this end, we employ the method of the hierarchy of correlations 
and compare different levels of accuracy.
To leading order, the usual free quasi-particle dynamics 
(as encoded in the two-point correlation functions) yields the 
standard picture of pre-thermalization.
Taking into account the back-reaction of these quasi-particle 
fluctuations onto the mean-field background as the first next-to-leading 
order effect, we observe a strong degradation of pre-thermalization, 
especially in low dimensions.
In contrast, the inclusion of three-point correlations enhances 
pre-thermalization.
\end{abstract}

\maketitle

\section{Introduction}

The question of how quantum many-body systems relax back to thermal 
equilibrium after being excited by an external stimulus has been the 
topic of active research \cite{D91,B93,S94,B03,C06,C07,R07,M07,R08,E09,M09,C10,P11,Ko11,Ki11,G11,B11,R12,G12,S14,St14,So14,Sm15,B15,C16,Ka16,L16,N16,F17,S17,N17,Hes17,Ma17,Her17,W17,H18,Fi18,Fr18,P18,La18,Wei19,R19,Pe21,B22,Al22,Ji23,Le23} and is still not completely understood -- 
especially for strongly interacting systems. 
As an indicator for the complexity of this problem, this relaxation 
dynamics can occur in various stages and on different time scales. 

To be more specific, let us consider a global stimulus in the form of 
a quench by suddenly (or rapidly) changing one or more parameters of 
the system -- which drives it out of equilibrium.
In a quasi-particle picture, this departure from equilibrium can be 
understood as the excitation of many quasi-particle modes. 
Moreover, such a quench typically corresponds to a coherent stimulus 
(rather than an incoherent, e.g., thermal, excitation) 
such that these quasi-particle modes all have the same phase -- 
at least initially. 
As a result of this coherence, local quantities typically start 
oscillating after the quench as the phases of the quasi-particle modes
evolve with time. 
However, due to the distribution or dispersion of the 
quasi-particle energies, their phases evolve differently and thus 
get scrambled.
As a result, the fluctuations of local quantities diminish gradually, approaching quasi-stationary values as time progresses.

Note, however, that this quasi-stationary state is not necessarily  
the thermal equilibrium state -- only the phases of the quasi-particle 
modes have become scrambled, their distribution has not changed yet 
and generally differs from a thermal distribution \cite{U20,M18}.
Thus, this initial stage of de-phasing is usually referred to as 
pre-thermalization -- full thermalization would also require the 
approach to thermal distribution functions, which can be achieved 
via Boltzmann collision dynamics (and requires interactions between 
the quasi-particles) \cite{M19,MR18,QS19,QSKS19,QSR23}.  

This complexity of the relaxation dynamics is already present for 
weakly interacting systems, but it can become even more challenging 
for strongly interacting quantum many-body systems. 
In the following, we study the pre-thermalization dynamics in the 
Mott insulator state of the Fermi-Hubbard model.
We place special emphasis on the impact of back-reaction and 
higher-order correlations in order to understand how these effects 
might change the usual picture.
In order to base our investigations on a well-defined analytical 
expansion in terms of a control parameter, we employ the method 
of the hierarchy of correlations \cite{Nav10}.


\section{Fermi-Hubard model}

To analyze the pre-thermalization dynamics of a prototypical strongly interacting quantum 
many-body system, we consider the fermionic Hubbard model. 
The Hamiltonian governing the system is given by ($\hbar=1$)
\begin{align}\label{extHubb}
\hat{H}=-\frac{1}{Z}\sum_{\mu\nu s}T_{\mu\nu}\hat{c}_{\mu s}^\dagger\hat{c}^{}_{\nu s}
+U \sum_\mu \hat{n}_\mu^\uparrow \hat{n}_\mu^\downarrow\,.
\end{align}
The fermionic creation and annihilation operators at sites $\mu$ and $\nu$, 
with spin $s\in{\uparrow,\downarrow}$, are denoted as 
$\hat{c}^\dagger_{\mu s}$ and $\hat{c}_{\nu s}$, respectively.
The hopping matrix $T_{\mu\nu}(t)$ is only nonvanishing for 
nearest neighbors where it adopts the value of the tunneling rate 
$T(t)$ which can depend on time.
The coordination number $Z$ represents the number of nearest neighbors 
for a given lattice site.
In the following, we shall consider the scenario that the system is initially 
prepared in a stationary state at half filling for $T=0$.
Then the tunneling rate is switched 
to a finite value $T>0$.
For simplicity we consider the a sudden change $T(t)=T\Theta(t)$, but the analysis can 
easily be generalized to other scenarios. 

\subsection{The hierarchy of correlations}

Apart from finite lattices (e.g., the Hubbard dimer), 
exact solutions of the fermionic Hubbard model~\eqref{extHubb}  
are only available in one spatial dimension \cite{E05} or in the limit 
of infinite dimensions \cite{G92}.
%
%
To derive approximate solutions in finite but
high dimensions, we employ a hierarchical method suitable for systems 
with large coordination numbers $Z$.
Hence, we partition the reduced density matrices into correlations between lattice sites and on-site density matrices $\hat{\rho}_\mu$.
For instance, the correlation between two sites is defined as  $\hat{\rho}^\mathrm{corr}_{\mu\nu}=\hat{\rho}_{\mu\nu}-\hat{\rho}_\mu\hat{\rho}_\nu$.  Similarly, correlations among three sites can be expressed as $\hat{\rho}^\mathrm{corr}_{\mu\nu\lambda}=\hat{\rho}_{\mu\nu\lambda}-\hat{\rho}_\mu\hat{\rho}_\nu\hat{\rho}_\lambda-\hat{\rho}^\mathrm{corr}_{\mu\nu}\hat{\rho}_\lambda-\hat{\rho}^\mathrm{corr}_{\mu\lambda}\hat{\rho}_\nu-\hat{\rho}^\mathrm{corr}_{\nu\lambda}\hat{\rho}_\mu$, and so forth. 

The on-site density matrix $\hat{\rho}_\mu$ and the two-site correlators 
$\hat{\rho}_{\mu\nu}^\mathrm{corr}$ follow evolution equations, which can be represented schematically as shown \cite{Nav10}
\begin{align}
i\partial_t \hat{\rho}_\mu
&=
F_1(\hat{\rho}_\mu,
\hat{\rho}_{\mu\nu}^\mathrm{corr})=\mathcal{O}(1)\,,
\\
i\partial_t \hat{\rho}_{\mu\nu}^\mathrm{corr}
&=
F_2(\hat{\rho}_\mu,\hat{\rho}_{\mu\nu}^\mathrm{corr},\hat{\rho}_{\mu\nu\lambda}^\mathrm{corr})=\mathcal{O}(1/Z)\,,\\
i\partial_t \hat{\rho}_{\mu\nu\lambda}^\mathrm{corr}
&=
F_3(\hat{\rho}_\mu,\hat{\rho}_{\mu\nu}^\mathrm{corr},\hat{\rho}_{\mu\nu\lambda}^\mathrm{corr},\hat{\rho}_{\mu\nu\lambda\kappa}^\mathrm{corr})=\mathcal{O}(1/Z^2)
\,.
\label{twosite}
\end{align}
Similar equations apply to higher-order correlations.
The specific forms of the nonlinear functions $F_n$ are determined by the exact von Neumann equation for the density matrix of the Hubbard 
model~\eqref{extHubb}.
Analyzing the evolution equations for the correlators reveals that the scaling hierarchy remains preserved over time when the initial values satisfy 
$\hat{\rho}_{\mu}=\mathcal{O}(1)$, 
$\hat{\rho}_{\mu\nu}^\mathrm{corr}=\mathcal{O}(1/Z)$, 
and so on \cite{Queiss14, Queiss23}.

Introducing the following operators proves to be advantageous when 
analyzing the evolution equation
\begin{align}
\hat c_{\mu s I}=\hat c_{\mu s}\hat n_{\mu\bar s}^I=
\left\{
\begin{array}{ccc}
 \hat c_{\mu s}(1-\hat n_{\mu\bar s}) & {\rm for} & I=0 \,,
 \\ 
 \hat c_{\mu s}\hat n_{\mu\bar s} & {\rm for} & I=1\,.
\end{array}
\right.
\end{align}
Here $\bar{s}$ denotes the opposite spin to $s$.
Intuitively, $\hat c_{\mu s I}$ annihilates a fermion with spin $s$ 
from a doubly occupied lattice site $\mu$ for $I=1$ but creates 
an empty lattice site $\mu$ for $I=0$ and is thus a precursor for 
doublon and holon quasi-particle operators.

The site-local quantities follow the evolution equation
\begin{align}\label{onsite}
&i\partial_t \langle\N{\mu}{s}{I} \N{\mu}{\s{s}}{J} \rangle\nonumber\\
&=\frac{(-1)^I}{Z}\sum_{\kappa,K}T_{\mu\kappa}
\left[\langle\Cd{\mu}{s}{J} \C{\kappa}{s}{K}\crangle-\langle\Cd{\kappa}{s}{K} \C{\mu}{s}{J}\crangle\right]\nonumber\\
&+\frac{(-1)^J}{Z}\sum_{\kappa,K}T_{\mu\kappa}
\left[\langle\Cd{\mu}{\s{s}}{I} \C{\kappa}{\s{s}}{K}\crangle-\langle\Cd{\kappa}{\s{s}}{K} \C{\mu}{\s{s}}{I}\crangle\right].
\end{align}
By definition, the 
two-site correlators between sites $\mu$ and $\nu$ are non-zero only 
for $\mu\neq\nu$.
This is expressed formally as $\langle\Cd{\mu}{s}{I} \C{\nu}{s}{J}\crangle=\langle\Cd{\mu}{s}{I} \C{\nu}{s}{J}\rangle-\delta_{\mu\nu}\delta_{IJ}  \langle\N{\mu}{s}{1} \N{\mu}{\s{s}}{I} \rangle$.
The behavior of these two-site correlators is determined by their 
evolution equation
\begin{align}\label{twositereal}
& i\partial_t\langle\Cd{\mu}{s}{I} \C{\nu}{s}{J}\crangle=U(J-I)\langle\Cd{\mu}{s}{I} \C{\nu}{s}{J}\crangle\nn
&+\sum_{\kappa,K}\frac{T_{\mu\kappa}}{Z}
\langle \N{\mu}{\s{s}}{I}\rangle \langle \Cd{\kappa}{s}{K} \C{\nu}{s}{J}\crangle
+\frac{T_{\mu\nu}}{Z}
\langle \N{\mu}{\s{s}}{I}\rangle\langle\N{\nu}{s}{1}\N{\nu}{\s{s}}{J}\rangle\nn
&-\sum_{\kappa,K}\frac{T_{\nu\kappa}}{Z}
\langle \N{\nu}{\s{s}}{J}\rangle \langle \Cd{\mu}{s}{I} \C{\kappa}{s}{K}\crangle-\frac{T_{\mu\nu}}{Z}\langle\N{\nu}{\s{s}}{J}\rangle\langle\N{\mu}{s}{1}\N{\mu}{\s{s}}{I}\rangle \nn
&-\delta_{\mu\nu}\sum_{\kappa,K}\frac{T_{\mu\kappa}}{Z}
\bigg[\langle \N{\mu}{\s{s}}{I}\rangle\langle\Cd{\kappa}{s}{K} \C{\mu}{s}{J}\crangle\nonumber\\
&-\langle \N{\mu}{\s{s}}{J}\rangle\langle\Cd{\mu}{s}{I} \C{\kappa}{s}{K}\crangle\bigg]+Q_{\mu\nu,s}^{IJ}\,.
\end{align}
The first two lines of equation (\ref{twositereal}) delineate the correlators' unhindered evolution and their interplay with site-local quantities.
The last two lines introduce terms designed to preserve the trace-free nature of the correlators, consequently fostering coupling among the modes, as detailed below.
Formally, this amounts to a minor correction of order $\mathcal{O}(1/Z^2)$ and can be safely disregarded to leading order.
The interactions with three-site correlations are encapsulated in $Q_{\mu\nu,s}^{IJ}$, also manifesting at the order of $\mathcal{O}(1/Z^2)$.
We have omitted considerations of particle-number correlations and spin-correlators.
Their dynamics unfold at a slower pace compared to the dynamics of doublon and holon excitations, hence assuming a subordinate role in the equilibration process.

\subsection{Normal state}

We consider the Hubbard system~\eqref{extHubb}
to be at half filling in the strong coupling limit with a large $U$.
%
%
In this limit, a small, finite temperature would not generate 
doublon-holon pairs but would tend to disrupt the spin order 
within the system.
%
%
This motivates the following ansatz for the site-local density matrix
\begin{align}
\hat{\rho}_\mu=&\left(\frac{1}{2}-\mathfrak{D}\right)\left(|\uparrow\rangle_\mu \langle \uparrow|+|\downarrow\rangle_\mu \langle \downarrow|\right)\nn
&+\mathfrak{D}\left(|\uparrow \downarrow\rangle_\mu \langle \uparrow \downarrow|+|0\rangle_\mu \langle 0|\right)\,.
\end{align}
Here, $\mathfrak{D}$ represents the double occupancy in the Hubbard 
system, which is zero before the quench dynamics.
If $T$ takes a finite value, correlations among lattice sites are 
generated, resulting in a non-zero double occupancy.
The dynamics can be determined from the local evolution equation (\ref{onsite}), which, in Fourier space, reads
\begin{align}\label{double}
i\partial_t \mathfrak{D}=\sum_{s}\int_\mathbf{k} T_\mathbf{k}
\left[f^{01}_{s}(T_\mathbf{k})-f^{10}_{s}(T_\mathbf{k})\right]\,. 
\end{align}
Here, the $f^{IJ}_{\mathbf{k},s}$ denote the Fourier components of 
the two-site correlators $\langle\Cd{\mu}{s}{I} \C{\nu}{s}{J}\crangle$ 
where we have assumed spatial homogeneity. 
Similarly, for the spatially homogeneous quench scenario 
under consideration, the $\mathbf{k}$-dependence solely stems from 
the Fourier transformation $T_\mathbf{k}$ of the hopping matrix
$T_{\mu\nu}$, which simplifies the momentum dependence 
$f^{IJ}_{\mathbf{k},s}=f^{IJ}_s(T_\mathbf{k})$ 
of the correlation functions (up to the accuracy we are interested in).

%
%
This permits the usage of the spectral function $\sigma_d(\omega)$.
For a hypercubic lattice in $d$ dimensions, we find
\begin{align}\label{spectral}
\sigma_d(\omega)&=\int_\mathbf{k}\delta(\omega-T_\mathbf{k})\nn
&=\frac{1}{2\pi}\int_{-\infty}^\infty dx\; e^{i x \omega }\left[\mathcal{J}_0\left(\frac{Tx}{d}\right)\right]^d\,,
\end{align}
where $\mathcal{J}_0$ denotes a Bessel function of the first kind.
Then, the relation (\ref{double}) takes the form
\begin{align}\label{double2}
i\partial_t \mathfrak{D}=\sum_s\int_{-T}^T d\omega\,\sigma_d(\omega)\, \omega \left[f^{01}_{ s}(\omega)-f^{10}_{ s}(\omega)\right]\,.
\end{align}
This simplifies the evaluation of a $d$-dimensional momentum integral in (\ref{double}) to the evaluation of a one-dimensional integral in (\ref{double2}).
Note that this simplification can only be used when neglecting spin correlations.
Similarly, the equations of motion for the two-site correlators can be written as
\begin{align}\label{corromega}
 (i \partial_t &+U^I-U^J)f^{IJ}_{  s}(\omega)=\frac{\omega}{2}\sum_L \left(f_{ s}^{LJ}(\omega)-f_{ s}^{IL}(\omega) \right)\nonumber\\
&-\frac{1}{2}\int_{-T}^{T} d\omega'\sigma_d(\omega')\,
\omega'\sum_L\left(f_{ s}^{LJ}(\omega')-f_{s}^{IL}(\omega') \right)\nn
&+\omega\left(\delta^{J1}-\delta^{I1}\right)
\left(\mathfrak{D}-\frac{1}{4}\right)+Q_{ s}^{IJ}(\omega)\,.
\end{align}
The first line of (\ref{corromega}) determines the free evolution of the 
individual modes.
The trace-free condition of the correlators leads to the coupling of the 
modes which is taken care of in the second line.
The primary source term of the correlators at leading order 
in the third line
arises from the lattice filling, with a minor correction introduced 
by the double occupancy.
Contributions from the interactions of three-point correlators are encapsulated in $Q_{s}^{IJ}(\omega)$.

\section{Quench dynamics}

In the upcoming discussion, we examine a quantum quench within the Mott regime, transitioning from $T=0$ to $T/U\ll 1$.
Given that the hierarchy of the equations of motion is effectively managed by a small expansion parameter $1/Z$, we can systematically include higher-order terms in our analysis.
Subsequently, we will delve into the analysis of pre-thermalization dynamics, employing various levels of approximation within this hierarchical expansion.

\subsection{Free quasi-particle evolution}

Considering first order in $1/Z$, we can disregard the coupling to the double occupancy and also omit 
the coupling among the modes in equation (\ref{corromega}), as they are formally of order $1/Z^2$, as shown in relation (\ref{twositereal}).
We can significantly simplify the analysis by transforming the correlators into a diagonal basis that corresponds to the doublon and holon excitations.
We achieve this by employing a Bogoliubov transformation $\mathfrak{f}^{\mathfrak{ab}}_{s\omega}=\sum_{IJ}O_\omega^{\mathfrak{a} I}O^{\mathfrak{b} J}_\omega f^{IJ}_{s\omega} $,
where the matrix is defined as
\begin{align}
O_{\omega}^{\mathfrak{a}I}=\begin{pmatrix}
                            \cos \varphi_\omega &\sin \varphi_\omega\\
                            -\sin \varphi_\omega & \cos \varphi_\omega
                           \end{pmatrix}\,,
\end{align}
along with the rotation angle given by
\begin{align}
\tan \varphi_\omega=\frac{\sqrt{\omega^2+U^2}+U}{\omega}\,. 
\end{align}
In the rotated frame, the free evolution of the correlators 
\eqref{corromega} is described as follows
\begin{align}
i\partial_t \mathfrak{f}_{ s}^{\mathfrak{ab}}(\omega)=\left[E_\omega^\mathfrak{b}-E_\omega^\mathfrak{a}\right]\left(\mathfrak{f}_{s}^{\mathfrak{ab}}(\omega)-\frac{1}{2}O_\omega^{\mathfrak{a}1}O_\omega^{\mathfrak{b}1}\right)\,.
\end{align}
Here, we introduce the quasi-particle energies of doublons and holons, given by \cite{Ede90,Belk95,Voj98,Ble22,Kung15,QSR23,Herr97}
\begin{align}\label{energies}
E^\pm_{\omega}=\frac{1}{2}\left(U-\omega\pm \sqrt{\omega^2+U^2}\right)\,.
\end{align}
The rapidly varying correlators $\mathfrak{f}^{+-}_s$ and $\mathfrak{f}^{-+}_s$ experience a rate of change of order $U$, while the quantities 
$\mathfrak{f}_s^{--}$ and $\mathfrak{f}_s^{++}$ do not display temporal evolution within the leading-order approach.
We identify $\mathfrak{f}_s^{--}$ and $\mathfrak{f}_s^{++}$ as the distribution functions of doublon and holon excitations in the Mott-Hubbard system.
The distribution functions exhibit nontrivial dynamics when accounting for the back-reaction of the correlators onto the mean field, as discussed in the subsequent section.
Certainly, it is worth noting that the  long-term dynamics, specifically the relaxation of these distribution functions towards a thermal equilibrium state, can be comprehended within the hierarchical expansion at order $\mathcal{O}(1/Z^3)$.
Within this framework, a set of Boltzmann equations for the distribution functions can be derived, as demonstrated in \cite{QS19,QSKS19}.
However, since the Boltzmann evolution occurs considerably after the pre-thermalization dynamics, we will not delve into the associated scattering processes here.

The rapidly fluctuating doublon-holon correlations, denoted as $\mathfrak{f}_s^{+-}=[\mathfrak{f}_s^{-+}]^*$, govern the temporal evolution of correlations among lattice sites.
This determines the equilibration of correlators among different lattice sites.
Following some algebraic manipulation, we determine that, at leading order, the temporal evolution of the two-site correlators can be expressed in closed form, namely
%
%
\begin{align}
&\langle \hat{c}^\dagger_{\mu s}\hat{c}_{\nu s}\rangle^\mathrm{corr}=\int_\mathbf{k}\frac{T_\mathbf{k}U}{2(T_\mathbf{k}^2+U^2)}\nonumber\\
&\times\left(1-\cos\left[\sqrt{U^2+T_{\mathbf{k}}^2}\,t\right]\right)e^{\mathbf{k}\cdot(\mathbf{x}_\mu-\mathbf{x}_\nu)}\,.
\end{align}
Hence, the dephasing of the individual modes leads to the equilibration of site-correlations, converging to a stationary value on the order of $\mathcal{O}(T/U)$.
Since the correlators influence the mean-field background through relation (\ref{double}), the probability of having a non-vanishing double occupancy becomes non-zero
\begin{align}\label{double1Z}
\mathfrak{D}(t)&=
\int_\mathbf{k}\frac{T^2_\mathbf{k}}{T_\mathbf{k}^2+U^2}\left(1-\cos\left[\sqrt{U^2+T_{\mathbf{k}}^2}\,t\right]\right)\nonumber\\
&=\int_{-T}^T d\omega\, \frac{\sigma_d(\omega)\omega^2}{U^2+\omega^2}\left(1-\cos\left[\sqrt{U^2+\omega^2}\,t\right]\right)\,. 
\end{align}
We have computed the oscillatory dynamics for two, three, and five dimensions, as shown in Fig.~\ref{plots1Z}.
To estimate the decay of the oscillations to the quasi-stationary value of order $\mathcal{O}(T^2/U^2)$, we perform an asymptotic expansion of the integral (\ref{double1Z}).
By employing the spectral density given in (\ref{spectral}), we deduce a power-law decay as follows
\begin{align}\label{longtime}
\lim_{t\rightarrow\infty}\mathfrak{D}(t)-\mathfrak{D}_\mathrm{asym}\approx \frac{T^2}{U^2}\left(\frac{\tau}{t}\right)^{d/2}f(t),
\end{align}
where $f(t)$ is a highly oscillating function with a magnitude of order unity, and $\tau=\mathcal{O}(U/T^2)$ denotes the time scale on which the oscillations decline.
Furthermore, we observe that the pre-thermalization dynamics in low dimensions differs from the results observed within the framework of 
dynamical mean-field theory (DMFT) 
calculations in the formal limit of infinite dimensions.
In DMFT, only a few oscillations are required until the 
pre-thermal state is approached, as shown in \cite{EK09,E10,T14}.
%
However, one should be careful when comparing these results 
\cite{EK09,E10,T14} directly to our findings. 
First, DMFT is usually based on a different scaling with 
coordination number $Z$, namely $1/\sqrt{Z}$ instead of $1/Z$
as in Eq.~\eqref{extHubb}. 
Second, the quench considered in \cite{EK09,E10,T14} is an interaction 
quench starting at $U=0$ instead of the hopping quench starting at 
$J=0$ considered here. 
As a consequence, the change of the system parameters is far more drastic 
and thus the excitation induced by the interaction quench much stronger.
Third, by effectively mapping to a single-site problem, important information regarding the spatial structures 
(e.g., the $\mathbf{k}$ dependence) is lost.
%
%
Nonetheless, equation (\ref{longtime}) reveals that, even within 
the leading order approach, the oscillations decay faster for higher 
dimensions than for lower dimensions, aligning with results from DMFT.

Note that pre-thermalization in the Bose-Hubbard model does 
also occur faster (see, e.g., \cite{Queiss14,So14}).
This can be partially attributed to the different dispersion relation 
of the Bose-Hubbard model, where the square root does also contain 
a contribution linear in the hopping strength $T_\mathbf{k}$.

\subsection{Back-reaction}

\begin{figure}
\includegraphics[width=\columnwidth]{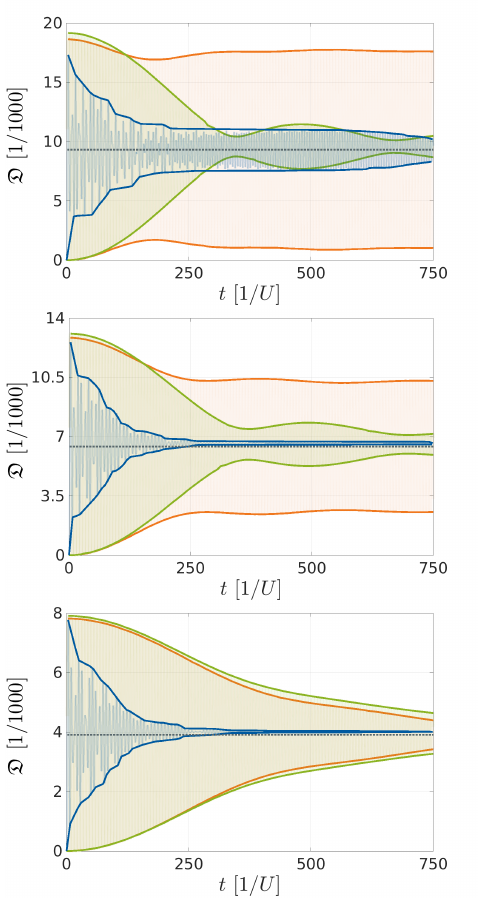}
\caption{Pre-thermalization dynamics of the double occupancy 
$\mathfrak{D}$ following a quench from $T=0$ to $T/U=0.2$ 
in two (top), three (middle), and five (bottom) dimensions.
%
The dynamics exhibit rapid oscillations; therefore, we have depicted only the envelope function illustrating the decay of the magnitude of the oscillations. The green curve represents the decay when only the free quasi-particle evolution is considered, as per Equation~\eqref{double1Z}. 
Adding the back-reaction of the correlations to the local mean field 
gives rise to the orange envelope. 
When considering three-point correlations in the dynamics, the blue 
envelope represents the decay of the correlators. 
The grey dotted line represents the asymptotic value 
$\mathfrak{D}_\mathrm{asym}$, as described in Equation~\eqref{Dasym}. 
}\label{plots1Z}
\end{figure}

The first departure from the free quasi-particle evolution involves the coupling to the double occupancy and inter-mode coupling.
By applying the Bogoliubov transformation from equation (\ref{corromega}) and using the particle-hole symmetry 
$f^{00}_{s}(\omega)=-f^{11}_{s}(-\omega)$, 
we derive the dynamics for the correlators
\begin{align}\label{eh}
i\partial_t \mathfrak{f}_{ s}^{\mathfrak{ab}}(\omega)=&\left[E_\omega^\mathfrak{b}-E_\omega^\mathfrak{a}\right]\left(\mathfrak{f}_{s}^{\mathfrak{ab}}(\omega)+2O_\omega^{\mathfrak{a}1}O_\omega^{\mathfrak{b}1}\left[\mathfrak{D}-\frac{1}{4}\right]\right)\nn
%
&+\frac{i}{4}\sum_I (-1)^I O^{\mathfrak{a}I}_\omega O^{\mathfrak{b}I}_\omega \partial_t \mathfrak{D}\,.
\end{align}
Each mode couples to the double occupancy. As the time-dependence of these forced oscillation is the same for each mode, there is a partial restoration of the coherences, altering the pre-thermalization process.
To quantify the correction to the leading-order analysis, 
we solve the system (\ref{eh}) along with (\ref{double2}) 
by utilizing Laplace transformations.
This yields $\mathfrak{D}(t)=\mathcal{L}^{-1}(\tilde{\mathfrak{D}}(r))$, where the Laplace transform of the double occupancy is given by
\begin{align}\label{doubleLaplace}
\tilde{\mathfrak{D}}(r)=\frac{1}{r}\frac{\mathfrak{I}(r)}{1+3\mathfrak{I}(r)}\,,
\end{align}
along with the integral
\begin{align}
 \mathfrak{I}(r)=\int_{-T}^T d\omega\, \sigma_d(\omega)\frac{\omega^2}{r^2+\omega^2+U^2}\,.
\end{align}
In the regime of strong interactions, the correction to the averaged value for the double occupancy (\ref{double1Z}) is relatively small, approximately on the order of $\mathcal{O}(T^4/U^4)$.
However, there are notable alterations in the dynamics, particularly 
in two and three dimensions, while the corrections in higher dimensions 
(greater than three) are comparatively minor.

To be more specific, let us first discuss the case of two 
dimensions (top panel of Fig.~\ref{plots1Z}).
Just considering the free quasi-particle evolution (green curves),
we see some beating effects, but a clear signature of pre-thermalization
(as expected). 
However, taking into account the back-reaction of these quasi-particle 
fluctuations onto the mean field (i.e., the double occupancy), 
we do not observe pre-thermalization anymore (orange curves), merely 
a slight initial decline of the magnitude of oscillations. 
As an intuitive picture, the coupling of the double occupancy to each 
mode restores coherences of local quantities, which weakens the effect 
of dephasing.

%
%
This effect of coherence restoration is also evident in three dimensions
(middle panel of Fig.~\ref{plots1Z}),
albeit with a somewhat stronger initial decay in magnitude.
Consequently, we conclude that especially in lower dimensions, 
it is crucial to consider higher-order correlators to accurately 
describe the pre-thermalization dynamics 
(blue curves in Fig.~\ref{plots1Z}).

In dimensions larger than three, the correction resulting from the 
back-reaction is significantly less pronounced. 
Surprisingly, it even amplifies the decay of oscillations compared 
to free evolution, as depicted in the lower panel of Fig.~\ref{plots1Z}.

We conclude that the accuracy of the hierarchical expansion for 
temporal evolution is highly contingent on spatial dimension and the 
chosen order of approximation.
%
%
Although in low dimensions higher-order correlators are essential 
for accurately describing the temporal evolution of local quantities, 
the asymptotic value can still be approximated if only two-site 
correlators and site-local dynamics are taken into account.

This observation holds true for the distribution functions 
representing doublon and holon excitations as well.
Considering that we commence the time-evolution in the Mott insulating state at $T=0$, the initial conditions dictate $\mathfrak{f}_s^{--}=\mathfrak{f}_s^{++}=0$.
Upon quenching the system to a finite value of $T/U$, these quasi-particle distributions also attain finite values. In fact, their pre-thermal values exhibit the same magnitude as the local double occupancy, namely
\begin{align}
\mathfrak{f}^{--}_{ s,\mathrm{asym}}(\omega)=\frac{U \mathfrak{D}_\mathrm{asym}}{4\sqrt{\omega^2+U^2}}=-\mathfrak{f}^{++}_{ s,\mathrm{asym}}(\omega)\,. 
\end{align}
These expressions, which are formally asymptotic, serve as the foundation for the distribution functions of doublons and holons, initiating the slow long-term dynamics.
During the pre-thermalization 
process, it is difficult to assert that 
$\mathfrak{f}^{--}_s$ and $\mathfrak{f}^{++}_s$ represent 
well-defined quasi-particle distributions, given their highly 
oscillatory nature.
Only as the temporal evolution approaches the pre-thermal state, 
do these distribution functions accurately describe slow variables. 
Therefore, the pre-thermalization process is essential to achieve the separation of time-scales. 
Initially, there is rapid oscillatory behavior, followed by a subsequent slow evolution, 
where a Boltzmann description becomes applicable.
In essence, by using the hierarchical equations of motion we do not assume the existence of quasi-particles beforehand.
Instead, the temporal evolution reveals the relevant slow-evolving variables directly \cite{QS19,QSKS19,Pi21}.
However, it is evident that for two and three spatial dimensions, a viable description of the equilibration process necessitates considering higher-order correlations.
As discussed below, even in this scenario, the average value of $\mathfrak{D}$ aligns with its asymptotic value.
Formally, the asymptotic value for the double occupancy can be straightforwardly deduced from the Laplace transform (\ref{doubleLaplace}), yielding
\begin{align}\label{Dasym}
\mathfrak{D}_\mathrm{asym}=\frac{\mathfrak{I}(0)}{1+3\mathfrak{I}(0)} \,.
\end{align}
Note that this expression is formally bounded by the infinite temperature value, i.e. $\mathfrak{D}_\mathrm{asym}<1/4$.
Comparing this with the leading-order result of $\mathfrak{D}_\mathrm{asym}$ obtained from (\ref{double1Z}), we observe that the correction due to the back-reaction is of order $\mathcal{O}(T^4/U^4)$.
As one might expect, the two-site correlations also experience a small correction,
\begin{align}
\langle \hat{c}^\dagger_{\mu s}\hat{c}_{\nu s}\rangle_\mathrm{asym}^\mathrm{corr}=\int_\mathbf{k}
\frac{UT_\mathbf{k}}{2(T_\mathbf{k}^2+U^2)}\frac{e^{\mathbf{k}\cdot(\mathbf{x}_\mu-\mathbf{x}_\nu)}}{1+3\mathfrak{I}(0)}\,.
\end{align}
%
We want to stress that even though we took the formal limit of $t\rightarrow \infty$ to derive the pre-thermal values, our analysis does not encompass the scenario of infinitely long times.
As mentioned earlier, to investigate the long-time dynamics, it is necessary to consider scattering processes between quasi-particles. These processes can be addressed using Boltzmann equations that involve four-site correlators, which are of order $\mathcal{O}(1/Z^3)$.

\subsection{Three-point correlators}

We emphasize the necessity of considering higher-order correlators to obtain a reasonable description of the pre-thermalization dynamics, especially in two and three dimensions.
The dominant three-site correlators, which contribute to the source term $Q_{\mu\nu,s}^{IJ}$ in equation (\ref{twositereal}), are 
given 
for pairwise distinct sites as follows
\begin{align}
&\langle \N{\lambda}{\s{s}}{K} \Cd{\mu}{s}{I}\C{\nu}{s}{J}\crangle=\langle \N{\alpha}{\s{s}}{K} \Cd{\mu}{s}{I}\C{\nu}{s}{J}\rangle-\langle \N{\alpha}{\s{s}}{K}\rangle \langle \Cd{\mu}{s}{I}\C{\nu}{s}{J}\rangle \label{threeone}\,,\\
&\langle \cd{\lambda}{s} \c{\lambda}{\s{a}}\Cd{\mu}{\s{s}}{I}\C{\nu}{s}{J}\crangle=\langle \cd{\lambda}{a} \c{\lambda}{\s{a}}\Cd{\mu}{\s{a}}{I}\C{\nu}{a}{J}\rangle\label{threetwo}\,,\\
&\langle\cd{\lambda}{s} \cd{\lambda}{\s{s}}\C{\mu}{\s{s}}{I}\C{\nu}{s}{J}\crangle=
\langle\cd{\lambda}{s} \cd{\lambda}{\s{s}}\C{\mu}{\s{s}}{I}\C{\nu}{s}{J}\rangle\,.\label{threethree}
\end{align}
Note that these simple expressions rely on the symmetries of our 
set-up, such as vanishing spin polarization which implies 
$\langle\C{\mu}{\s{s}}{I}\C{\nu}{s}{J}\rangle=0$.

These correlators are described by equations of motion that involve couplings to two-site correlations, double occupancy, and higher-order correlators (see Appendix \ref{threepoint} for details).
The resulting equations of motion are highly nonlinear, disrupting coherences among individual modes and driving local quantities towards a long-lived pre-thermal state.
In dimensions two and three, the effect of coherence restoration due to the back-reaction is surpassed, leading to a significant decrease in the magnitude of the oscillations, as illustrated in Fig.~\ref{plots1Z}.

The site-local dynamics are driven by hopping events of holon and doublon excitations. As the return probability to a particular site diminishes with higher dimensionality, coherent oscillations decay more rapidly in higher dimensions. This observation is consistent with our results, where after incorporating the three-point correlators, the equilibration process accelerates in dimensions three and five compared to the two-dimensional setting.

\section{Conclusions}

Via the hierarchy of correlations, we have investigated the 
pre-thermalization dynamics in the  Mott insulator 
state of the strongly interacting Fermi-Hubbard model 
after a hopping quench. 
As a starting point, we focused on the free quasi-particle evolution 
as encoded in the two-point correlation functions which yields the 
usual pre-thermalization picture. 
However, even in three dimensions we find that pre-thermalization
takes a comparably long time $\Delta t\gg1/U$, i.e., much longer than 
what is expected from the limit of infinite dimensions.

In view of this comparably long time scale, small corrections to this 
leading order might become important as they could accumulate over time. 
Taking into account the first non-trivial correction to this leading 
order, i.e., the back-reaction of the quasi-particle fluctuations onto 
the mean-field background, we find that this effect significantly 
suppresses pre-thermalization, especially in lower dimensions. 
As an intuitive picture, the joint coupling of all the quasi-particle 
modes to the same mean-field mode introduces additional coherences 
between them and reduces their de-phasing. 

We also 
included three-point correlations (i.e., further higher-order effects)
in our approach.  
Their impact tends to enhance pre-thermalization 
(even stronger than the suppression due to back-reaction), 
which can be explained by the fact that they mediate non-linear 
interactions between the quasi-particle modes -- which in turn 
can result in a more efficient scrambling of their phases. 

Note, however that considering these three-point correlations is not 
sufficient for describing full thermalization. 
This occurs on much longer time scales on which the quasi-particle
distribution functions change via Boltzmann type collisions.
Describing them requires incorporating the four-point correlations,
see also \cite{QS19,QSKS19,QSR23}.

\bigskip 

\acknowledgments 

Funded by the Deutsche Forschungsgemeinschaft 
(DFG, German Research Foundation) -- Project-ID 278162697-- SFB 1242. 


\appendix
%
%
%

\begin{widetext}
\section{Three-point correlators}\label{threepoint}
%
%
We consider only the particular case when the momentum dependence of 
Fourier components of the correlators is solely determined by the hopping matrix $T_\mathbf{k}$.
Then we can employ for the correlators (\ref{threeone})-(\ref{threethree}) the expansions 
\begin{align}
\langle \N{\lambda}{\s{s}}{K} \Cd{\mu}{s}{I}\C{\nu}{s}{J}\crangle&=\int_{\mathbf{k},\mathbf{p}}
g_{\bar{s}ss}^{KIJ}(T_\mathbf{k},T_\mathbf{p})e^{i\mathbf{k}\cdot (\mathbf{x}_\mu-\mathbf{x}_\lambda)+i\mathbf{p}\cdot (\mathbf{x}_\nu-\mathbf{x}_\lambda)}\,,\\
\langle \cd{\lambda}{s} \c{\lambda}{\s{a}}\Cd{\mu}{\s{s}}{I}\C{\nu}{s}{J}\crangle&=\int_{\mathbf{k},\mathbf{p}} r_{\bar{s}s}^{IJ}(T_\mathbf{k},T_\mathbf{p})e^{i\mathbf{k}\cdot (\mathbf{x}_\mu-\mathbf{x}_\lambda)+i\mathbf{p}\cdot (\mathbf{x}_\nu-\mathbf{x}_\lambda)}\,,
\\
\langle\cd{\lambda}{s} \cd{\lambda}{\s{s}}\C{\mu}{\s{s}}{I}\C{\nu}{s}{J}\crangle&=
\int_{\mathbf{k},\mathbf{p}}h_{\bar{s}s}^{IJ}(T_\mathbf{k},T_\mathbf{p})e^{i\mathbf{k}\cdot (\mathbf{x}_\mu-\mathbf{x}_\lambda)+i\mathbf{p}\cdot (\mathbf{x}_\nu-\mathbf{x}_\lambda)}\,.
\end{align}
Employing the spectral density, we obtain for the source term in (\ref{corromega})
the expression
\begin{align}
Q^{IJ}_s(\omega)&=\sum_K\int_{-T}^T d\omega' \sigma_d(\omega')\omega'\left[ g^{IKJ}_{\s{s}ss}(\omega',\omega)-g^{JIK}_{\s{s}ss}(\omega,\omega')\right]\nn
&+(-1)^I \sum_K\int_{-T}^T d\omega' \sigma_d(\omega')\omega'\left[r_{\s{s}s}^{KJ}(\omega',\omega)+h^{KJ}_{\s{s}s}(\omega',\omega)\right]\nn
&-(-1)^J  \sum_K\int_{-T}^T d\omega' \sigma_d(\omega')\omega'\left[\left(r^{KI}_{\s{s}s}(\omega',\omega)\right)^*+\left(h^{KI}_{\s{s}s}(\omega,\omega')\right)^* \right]\,.
\end{align}
For the first set of three-point-correlators, the equation of motion reads
\begin{align}
(i\partial_t +U^I-U^J)g^{KIJ}(\omega_1,\omega_2)&=\frac{\omega_1}{2}\sum_L g^{KLJ}_{\s{s}ss}(\omega_1,\omega_2)-
\frac{\omega_2}{2}\sum_L g^{KIL}_{\s{s}ss}(\omega_1,\omega_2)\nn
&+\frac{(-1)^K}{4}\omega_1 \left[f^{0J}_s(\omega_2)-f^{1J}_s(\omega_2)\right]-
\frac{(-1)^K}{4}\omega_2 \left[f^{I0}_s(\omega_1)-f^{I1}_s(\omega_1)\right]\nn
&-\frac{1}{2}\int_{-T}^T d\omega\,\sigma_d(\omega)\omega\sum_L\left[g^{KLJ}_{\s{s}ss}(\omega,\omega_2)-g_{\s{s}ss}^{KIL}(\omega_1,\omega)\right]\,.
\end{align}
The last line ensures the sum rules $\int d\omega\, \sigma_d(\omega)g^{KIJ}_{\s{s}ss}(\omega,\omega_2)=\int d\omega\, \sigma_d(\omega)g^{KIJ}_{\s{s}ss}(\omega_1,\omega)=0 $ which follow from the requirement that the three-point correlators have to vanish if two sites coincide.
The equation of motion for the second set of three-point correlators contains bilinear 
couplings among the two-site correlators.
It reads explicitely
\begin{align}
(i\partial_t +U^I-U^J)r^{IJ}_{\s{s}s}(\omega_1,\omega_2)&= \frac{\omega_1}{2}\sum_L r^{LJ}_{\s{s}s}(\omega_1,\omega_2)-
\frac{\omega_2}{2}\sum_L r^{IL}_{\s{s}s}(\omega_1,\omega_2)\nn
&+\sum_{K,L}(\omega_1-\omega_2) f_s^{KJ}(\omega_2)f_{\s{s}}^{IL}(\omega_1)\nn
&-\sum_L \left[(-1)^I\left(\mathfrak{D}-\frac{1}{4}\right)-\frac{1}{4}(-1)^L\right]\omega_1 f^{LJ}_s(\omega_2)\nn
&
+\sum_L \left[(-1)^J\left(\mathfrak{D}-\frac{1}{4}\right)-\frac{1}{4}(-1)^L\right] \omega_2 f^{IL}_{\s{s}}(\omega_1)\nn
&-\frac{1}{2}\int_{-T}^T d\omega \sigma_d(\omega)\, \omega\sum_L \left[r^{LJ}_{\s{s}s}(\omega,\omega_2)-r^{IL}_{\s{s}s}(\omega_1,\omega)\right]\nn
&-\int_{-T}^T d\omega \sigma_d(\omega)\, \omega \sum_{K,L}\left[f_s^{KJ}(\omega_2)f_{\s{s}}^{IL}(\omega)-f_s^{KJ}(\omega)f_{\s{s}}^{IL}(\omega_1)\right]\,.
\end{align}
Again, the last two lines ensure the sum rules $\int d\omega\, \sigma_d(\omega)r^{IJ}_{\s{s}s}(\omega,\omega_2)=\int d\omega\, \sigma_d(\omega)r^{IJ}_{\s{s}s}(\omega_1,\omega)=0$.
Finally, we have for the third set of equations
\begin{align}
(i\partial_t -U^I-U^J+U)h^{IJ}(\omega_1,\omega_2)&=-\frac{\omega_1}{2}\sum_L h^{LJ}(\omega_1,\omega_2)
-\frac{\omega_2}{2}\sum_L h^{IL}(\omega_1,\omega_2)\nn
&+\sum_{K,L}(\omega_1+\omega_2)f^{KI}_{\s{s}}(\omega_1)f^{LJ}_s(\omega_2)\nn
&+\omega_1 \sum_L\left[\frac{(-1)^L}{4}+(-1)^I\left(\frac{1}{4}-\mathfrak{D}\right)\right]f^{LJ}_s(\omega_2)\nn
&+\omega_2 \sum_L\left[\frac{(-1)^L}{4}+(-1)^J\left(\frac{1}{4}-\mathfrak{D}\right)\right]f^{LI}_{\s{s}}(\omega_1)\nn
&+\frac{1}{2}\int_{-T}^T d\omega \sigma_d(\omega)\, \omega\sum_L \left[h_{\s{s}s}^{LJ}(\omega,\omega_2)+h_{\s{s}s}^{IL}(\omega_1,\omega)\right]\nn
&-\int_{-T}^T d\omega \sigma_d(\omega)\, \omega\sum_{K,L}\left[f_{\s{s}}^{KI}(\omega_1)f_s^{LJ}(\omega)+f_{\s{s}}^{KI}(\omega)f_s^{LJ}(\omega_2)\right]\,.
\end{align}
Also here, the last two lines ensure the sum rules $\int d\omega\, \sigma_d(\omega)h^{IJ}_{\s{s}s}(\omega,\omega_2)=\int d\omega\, \sigma_d(\omega)h^{IJ}_{\s{s}s}(\omega_1,\omega)=0$.

\end{widetext}

\end{document}